\documentclass[aps,prd,preprint,superscriptaddress]{revtex4}

\usepackage{graphicx}

\def\ie{{\it i.e.}}
\def\eg{{\it e.g.}}

\def\ld{\lambda}

\def\nn{\nonumber}

\def\bwt{\begin{widetext}}
\def\ewt{\end{widetext}}
\def\be{\begin{equation}}
\def\ee{\end{equation}}
\def\bea{\begin{eqnarray}} 
\def\eea{\end{eqnarray}}
\def\bean{\begin{eqnarray*}}
\def\eean{\end{eqnarray*}}
\def\bary{\begin{array}}
\def\eary{\end{array}}
\def\bit{\begin{itemize}}
\def\eit{\end{itemize}}

\def\ld{\lambda}

\def\su5u1{SU(5) \times U(1)}
\def\fsu5u1{SU(5) \times U(1)'}
\def\so10{SO(10)}
\def\sq20{SO(10) \times SO(10)}
\def\sla{/\!\!\!\!A_\mu}
\def\slb{/\!\!\!\!B_\mu}

\begin{document}

\begin{flushright}
ANL-HEP-PR-04-66\\
hep-ph/0407180\\
\end{flushright}

\title{Supersymmetric $SO(10)$ Models Inspired by Deconstruction}
\author{Chao-Shang Huang}
\affiliation{Institute of Theoretical Physics, Academia Sinica,
  Beijing 100080, China}

\author{Jing Jiang}
\affiliation{High Energy Physics Division, Argonne National
  Laboratory, Argonne, IL 60439}

\author{Tianjun Li}
\affiliation{School of Natural Science, Institute for Advanced Study,
  Einstein Drive, Princeton, NJ 08540}

\date{\today}

\begin{abstract}
  We consider 4-dimensional $N=1$ supersymmetric $SO(10)$ models
  inspired by deconstruction of 5-dimensional $N=1$ supersymmetric
  orbifold $SO(10)$ models and high dimensional non-supersymmetric
  $SO(10)$ models with Wilson line gauge symmetry breaking. We discuss
  the $SO(10) \times SO(10)$ models with bi-fundamental link fields
  where the gauge symmetry can be broken down to the Pati-Salam,
  $SU(5)\times U(1)$, flipped $SU(5)\times U(1)'$ or the standard
  model like gauge symmetry.  We also propose an $SO(10)\times
  SO(6)\times SO(4)$ model with bi-fundamental link fields where the
  gauge symmetry is broken down to the Pati-Salam gauge symmetry, and
  an $SO(10)\times SO(10)$ model with bi-spinor link fields where the
  gauge symmetry is broken down to the flipped $SU(5)\times U(1)'$
  gauge symmetry.  In these two models, the Pati-Salam and flipped
  $SU(5)\times U(1)'$ gauge symmetry can be further broken down to the
  standard model gauge symmetry, the doublet-triplet splittings can be
  obtained by the missing partner mechanism, and the proton decay
  problem can be solved.  We also study the gauge coupling
  unification. We briefly comment on the interesting variation models
  with gauge groups $SO(10)\times SO(6)$ and $SO(10)\times {\rm
    flipped}~ SU(5)\times U(1)'$ in which the proton decay problem can
  be solved.

\end{abstract}

\maketitle


\section{Introduction}
Supersymmetry (SUSY) provides an elegant solution to the gauge
hierarchy problem, and grand unified theories (GUTs) give us a simple
understanding of the quantum numbers of the standard model (SM)
fermions. In addition, the success of gauge coupling unification in
the minimal supersymmetric standard model (MSSM) strongly supports the
possibility of the SUSY GUT.  Other appealing features include that
the electroweak symmetry breaking is induced by radiative corrections
due to the large top quark Yukawa coupling, and that tiny neutrino
masses can be naturally explained by the see-saw mechanism.
Therefore, SUSY GUT is one of the most promising candidates that
describe the known fundamental interactions in nature except gravity.
However, there are problems in the 4-dimensional SUSY GUT: the
doublet-triplet splitting problem, the proton decay problem, the
fermion mass problem, and the GUT gauge symmetry breaking mechanism.

During the last few years, orbifold GUTs have been studied
extensively~\cite{kawa, GAFF, LHYN, AHJMR, LTJ1, Hall:2001tn,
  Haba:2001ci, Asaka:2001eh, LTJ2, Dermisek:2001hp, Huang:2001bm,
  Li:2001tx, Barr:2002fb, Kyae:2002hu, Kim:2002im, Gogoladze:2003ci,
  Li:2003ee}.  The main point is that the supersymmetric GUT models
exist in 5 or 6 dimensional space-time, and they are broken down to
4-dimensional $N=1$ supersymmetric SM like models for the zero modes
through compactification on various orbifolds.  This is because the
orbifold parity projects out the zero modes of some components in
vector multiplet and hypermultiplets.  The GUT gauge symmetry breaking
problem, the doublet-triplet splitting problem and the fermion mass
problem have been solved elegantly by orbifold projections. The proton
decay problem can be solved because we can define a continuous
$U(1)_R$ symmetry.  Other interesting phenomenology, like flavour
symmetry from the $R$ symmetry, gauge-fermion unification, gauge-Higgs
unification, and gauge-Yukawa unification, have also been
studied~\cite{kawa, GAFF, LHYN, AHJMR, LTJ1, Hall:2001tn, Haba:2001ci,
  Asaka:2001eh, LTJ2, Dermisek:2001hp, Huang:2001bm, Li:2001tx,
  Barr:2002fb, Kyae:2002hu, Kim:2002im, Gogoladze:2003ci, Li:2003ee}.

On the other hand, deconstruction was proposed about three year ago to
latticize the gauge theories in higher
dimensions~\cite{Arkani-Hamed:2001ca, Hill:2000mu, Cheng:2001vd}.  The
idea of deconstruction is interesting because it gives a UV completion
to the higher dimensional theories.  Applying this idea to orbifold
SUSY GUTs, we are able to construct interesting 4-dimensional SUSY
GUTs where the problems in the usual 4-dimensional models can be
solved, in other words, the merits of orbifold GUTs can be
preserved~\cite{Csaki:2001qm, Cheng:2001qp, Li:2002xd}.  In addition,
inspired by deconstruction of orbifold SUSY GUTs, we can construct new
models which can not be obtained from orbifold GUTs while still retain
the nice properties in orbifold models.  Deconstruction of the
orbifold $SU(5)$ models were discussed in Ref.~\cite{Csaki:2001qm,
Cheng:2001qp, Li:2002xd}.  In these models the doublet-triplet
splitting problem and the proton decay problem can be solved.
Deconstruction may also provide insight into fermion masses and
mixings~\cite{Csaki:2001qm}. Since the number of fields in these
models is finite, the corrections to gauge couplings can be reliably
calculated, and there exist threshold corrections to the differential
runnings of the gauge couplings~\cite{Csaki:2001qm, Cheng:2001qp}.
Elements of deconstruction can be found in earlier
papers~\cite{Halpern:1975yj}.

In this paper, we construct 4-dimensional $N=1$ supersymmetric
$SO(10)$ models inspired by deconstruction of the 5-dimensional $N=1$
supersymmetric orbifold $SO(10)$ models and high dimensional
non-supersymmetric $SO(10)$ models with Wilson line gauge symmetry
breaking. We study $SO(10) \times SO(10)$ models with bi-fundamental
link fields where the gauge symmetry can be broken down to the
Pati-Salam (PS), $SU(5)\times U(1)$, flipped $SU(5)\times U(1)'$ or
the SM like gauge symmetry.  However, we need to fine-tune the
superpotential in the models, in order to have the doublet-triplet
splitting.  In addition, we propose an $SO(10)\times SO(6)\times
SO(4)$ model where the gauge symmetry can be broken down to PS gauge
symmetry, and an $SO(10)\times SO(10)$ model with bi-spinor link
fields (${\bf 16, {\overline{16}}}$) and (${\bf {\overline{16}}, 16
}$), in which the gauge symmetry is broken down to the flipped
$\fsu5u1$ gauge symmetry.  In these models, the gauge symmetry can be
further broken down to the SM gauge symmetry, and the doublet-triplet
splitting is naturally realized through the missing partner mechanism.
The proton decay due to dimension-5 operators is thus negligible and
the proton lifetime due to dimension-6 operators is well above the
current experimental bound because the GUT scale is at least a few
times $10^{16}$ GeV. Therefore, there is no proton decay problem.  We
also discuss the gauge coupling unification in these two models.
Furthermore, we briefly comment on the interesting variation models
with gauge groups $SO(10)\times SO(6)$ and $SO(10)\times {\rm
  flipped}~ SU(5)\times U(1)' $ where the proton decay problem can be
solved, and the $SO(10)\times SO(10)$ model with bi-spinor link fields
in which the gauge symmetry is broken down to the $SU(5)\times U(1)$
gauge symmetry.

We first give a brief review of the orbifold SUSY GUTs and the
non-SUSY GUTs with Wilson line gauge symmetry breaking, and
deconstruction of both types of models in Section~\ref{sec:review}.
We then discuss the models where the gauge symmetry can be broken down
to the diagonal PS gauge symmetry in Section~\ref{sec:patsal}. We
comment on the models inspired by deconstructions of non-SUSY GUTs
with Wilson line gauge symmetry breaking in Section~\ref{sec:noncom}.
In Section~\ref{sec:bispinor}, we consider gauge symmetry breaking
with bi-spinor link fields.  We summarize our results in
Section~\ref{sec:conclu}.

\section{Brief review of High Dimensional GUT breaking and deconstruction}
\label{sec:review}
\subsection{5-Dimensional Orbifold Supersymmetric GUTs}
In the 5-dimensional orbifold SUSY GUTs, the 5-dimensional manifold is
factorized into the product of ordinary 4-dimensional Minkowski
space-time $M^4$ and the orbifold $S^1/(Z_2\times Z'_2)$.  The
corresponding coordinates are $x^{\mu}$ ($\mu = 0, 1, 2, 3$) and
$y\equiv x^5$. The radius for the fifth dimension is $R$.  The
orbifold $S^1/(Z_2\times Z'_2)$ is obtained by $S^1$ moduloing the
equivalent class
\begin{eqnarray}
P:~~y \sim -y~,~P':~~ y' \sim -y'~,~\,
\end{eqnarray}
where $y'\equiv y-\pi R/2$.  There are two fixed points, $y=0$ and
$y=\pi R/2$.

The $N=1$ supersymmetric theory in 5-dimension have 8 real
supercharges, corresponding to $N=2$ supersymmetry in 4-dimension.  In
terms of the physical degrees of freedom, the vector multiplet
contains a vector boson $A_M$ with $M=0, 1, 2, 3, 5$, two Weyl
gauginos $\lambda_{1,2}$, and a real scalar $\sigma$.  In the
4-dimensional $N=1$ supersymmetry language, it contains a vector
multiplet $V \equiv (A_{\mu}, \lambda_1)$ and a chiral multiplet
$\Sigma \equiv ((\sigma+iA_5)/\sqrt 2, \lambda_2)$ which transform in
the adjoint representation of group $G$.  The 5-dimensional
hypermultiplet consists of two complex scalars $\phi$ and $\phi^c$,
and a Dirac fermion $\Psi$.  It can be decomposed into two chiral
mupltiplets $\Phi(\phi, \psi \equiv \Psi_R)$ and $\Phi^c(\phi^c,
\psi^c \equiv \Psi_L)$, which are in the conjugate representations of
each other under the gauge group.

The general action for the group $G$ gauge fields and their
couplings to the bulk hypermultiplet $\Phi$ is~\cite{NAHGW}
\begin{eqnarray}
S&=&\int{d^5x}\frac{1}{k g^2}
{\rm Tr}\left[\frac{1}{4}\int{d^2\theta} \left(W^\alpha W_\alpha+{\rm H.
C.}\right)
\right.\nonumber\\&&\left.
+\int{d^4\theta}\left((\sqrt{2}\partial_5+ {\bar \Sigma })
e^{-V}(-\sqrt{2}\partial_5+\Sigma )e^V+
\partial_5 e^{-V}\partial_5 e^V\right)\right]
\nonumber\\&&
+\int{d^5x} \left[ \int{d^4\theta} \left( {\Phi}^c e^V {\bar \Phi}^c +
 {\bar \Phi} e^{-V} \Phi \right)
\right.\nonumber\\&&\left.
+ \int{d^2\theta} \left( {\Phi}^c (\partial_5 -{1\over {\sqrt 2}} \Sigma)
\Phi + {\rm H. C.}
\right)\right]~.~\,
\end{eqnarray}

Under the parity operator $P$, the vector multiplet transforms as
\begin{eqnarray}
V(x^{\mu},y)&\to  V(x^{\mu},-y) = P V(x^{\mu}, y) P^{-1}
~,~\,
\end{eqnarray}
\begin{eqnarray}
 \Sigma(x^{\mu},y) &\to\Sigma(x^{\mu},-y) = - P \Sigma(x^{\mu}, y) P^{-1}
~.~\,
\end{eqnarray}
For the hypermultiplet $\Phi$ and $\Phi^c$, we have
\begin{eqnarray}
\Phi(x^{\mu},y)&\to \Phi(x^{\mu}, -y)  = \eta_{\Phi} P^{l_\Phi} \Phi(x^{\mu},y)
(P^{-1})^{m_\Phi}~,~\,
\end{eqnarray}
\begin{eqnarray}
\Phi^c(x^{\mu},y) &\to \Phi^c(x^{\mu}, -y)  = -\eta_{\Phi} P^{l_\Phi}
\Phi^c(x^{\mu},y)^{m_\Phi}
~,~\,
\end{eqnarray}
where $\eta_{\Phi}$ is $\pm$, $l_{\Phi}$ and $m_{\Phi}$ are
respectively the numbers of the fundamental index and anti-fundamental
index for the bulk multiplet $\Phi$ under the bulk gauge group $G$.
For example, if $G$ is an $SU(N)$ group, for fundamental
representation, $l_{\Phi}=1$, $m_{\Phi}=0$, and for adjoint
representation, $l_{\Phi}=1$, $m_{\Phi}=1$.  Moreover, the
transformation properties for the vector multiplet and hypermultiplets
under $P'$ are the same as those under $P$.

For $G=SU(5)$, to break the $SU(5)$ gauge symmetry, we choose the
following $5\times 5$ matrix representations for the parity operators
$P$ and $P'$
\begin{equation}
\label{eq:pppsu5}
P={\rm diag}(+1, +1, +1, +1, +1)~,~P'={\rm diag}(+1, +1, +1, -1, -1)
 ~.~\,
\end{equation}
Under the $P'$ parity, the gauge generators $T^\alpha$ ($\alpha =1$,
$2$, ..., $24$) for $SU(5)$ are separated into two sets: $T^a$ are the
generators for the SM gauge group, and $T^{\hat a}$ are the generators
for the broken gauge group
\begin{equation}
P~T^a~P^{-1}= T^a ~,~ P~T^{\hat a}~P^{-1}= T^{\hat a}
~,~\,
\end{equation}
\begin{equation}
P'~T^a~P^{'-1}= T^a ~,~ P'~T^{\hat a}~P^{'-1}= - T^{\hat a}
~.~\,
\end{equation}
The zero modes of the $SU(5)/SM$ gauge bosons are projected out, thus,
the 5-dimensional $N=1$ supersymmetric $SU(5)$ gauge symmetry is
broken down to the 4-dimensional $N=1$ supersymmetric SM gauge
symmetry for the zero modes. For the zero modes and KK modes, the
4-dimensional $N=1$ supersymmetry is preserved on the 3-branes at the
fixed points, and only the SM gauge symmetry is preserved on the
3-brane at $y=\pi R/2$.

For $G=SO(10)$, the generators $T^\alpha$ of $SO(10)$ are imaginary
antisymmetric $10 \times 10$ matrices.  In terms of the $2\times 2$
identity matrix $\sigma_0$ and the Pauli matrices $\sigma_i$, they can
be written as tensor products of $2\times 2$ and $5 \times 5$
matrices, $(\sigma_0, \sigma_1, \sigma_3) \otimes A_5$ and $\sigma_2
\otimes S_5$ as a complete set, where $A_5$ and $S_5$ are the $5\times
5$ real anti-symmetric and symmetric matrices.  The generators of the
$\su5u1$ are
\bea
&& \sigma_0 \otimes A_3\,, \quad \sigma_0 \otimes A_2\,, \quad \sigma_0
\otimes A_X \nonumber \\
&& \sigma_2 \otimes S_3\,, \quad \sigma_2 \otimes S_2\,, \quad
\sigma_2 \otimes S_X \,,
\eea
the generators for flipped $\fsu5u1$ are
\bea
&& \sigma_0 \otimes A_3\,, \quad \sigma_0 \otimes A_2\,, \quad
\sigma_1 \otimes A_X \nonumber \\
&& \sigma_2 \otimes S_3\,, \quad \sigma_2 \otimes S_2\,, \quad
\sigma_3 \otimes A_X \,,
\eea
and  the generators for
 $SU(4)_C \times SU(2)_L \times SU(2)_R$ are
\begin{equation}
\begin{array}{cc}
  (\sigma_0,\sigma_1,\sigma_3) \otimes A_3~,~ &
     (\sigma_0,\sigma_1,\sigma_3) \otimes A_2~,~  \\
  \sigma_2 \otimes S_3~,~ & \sigma_2 \otimes S_2~,~\,
\end{array}
\label{eq:422gen}
\end{equation}
where $A_3$ and $S_3$ are respectively the diagonal blocks of $A_5$
and $S_5$ that have indices 1, 2, and 3, while the diagonal blocks
$A_2$ and $S_2$ have indices 4 and 5. $A_X$ and $S_X$ are the off
diagonal blocks of $A_5$ and $S_5$.

We choose the $10\times 10 $ matrix  for $P$ as
\begin{eqnarray}
P&=& \sigma_0 \otimes {\rm diag}(1,1,1,1,1)~.~\,
\end{eqnarray}
To break the $SO(10)$ down to $SU(5)\times U(1)$, we choose
\begin{eqnarray}
\label{eq:psu5}
P'&=& \sigma_2 \otimes {\rm diag}(1,1,1,1,1)~,~\,
\end{eqnarray}
to break the $SO(10)$ down to flipped $\fsu5u1$, we choose
\begin{eqnarray}
\label{eq:pfsu5}
P'&=& \sigma_2 \otimes {\rm diag}(1,1,1,-1,-1)~,~\,
\end{eqnarray}
and to break the $SO(10)$ down to the PS gauge symmetry, we choose
\begin{eqnarray}
\label{eq:pps}
P'&=& \sigma_0 \otimes {\rm diag}(1,1,1,-1,-1)~.~\,
\end{eqnarray}
For the zero modes, the 5-dimensional $N=1$ supersymmetric $SO(10)$
gauge symmetry is broken down to the 4-dimensional $N=1$
supersymmetric $SU(5)\times U(1)$, flipped $\fsu5u1$ and the PS gauge
symmetries.  Including the KK modes, the 3-branes at the fixed points
preserve the 4-dimensional $N=1$ supersymmetry, and the gauge symmetry
on the 3-brane at $y=\pi R/2$ is $SU(5)\times U(1)$, flipped $\fsu5u1$
and the PS gauge symmetries, for different choices of $P'$.

We can also break the $SO(10)$ down to the SM like
($SU(3)_C\times SU(2)_L \times U(1)_Y \times U(1)_X$) gauge symmetry,
by choosing the following matrix representations for $P$ and $P'$
\begin{eqnarray}
\label{eq:psmx1}
P ~=~ \sigma_0 \otimes {\rm diag}(1,1,1,-1,-1) ~,~
P'~=~ \sigma_2 \otimes {\rm diag}(1,1,1,1,1)~,~\,
\end{eqnarray}
or
\begin{eqnarray}
\label{eq:psmx2}
P ~=~ \sigma_0 \otimes {\rm diag}(1,1,1,-1,-1) ~,~
P'~=~ \sigma_2 \otimes {\rm diag}(1,1,1,-1,-1)~,~\,
\end{eqnarray}
or
\begin{eqnarray}
\label{eq:psmx3}
P~=~ \sigma_2 \otimes {\rm diag}(1,1,1,1,1)~,~
P'~=~ \sigma_2 \otimes {\rm diag}(1,1,1,-1,-1)~.~\,
\end{eqnarray}

\subsection{High Dimensional non-Supersymmetric GUTs with Wilson Line Gauge
  Symmetry Breaking} Other than the orbifold GUTs, Wilson line gauge
symmetry breaking can be applied to break high dimensional gauge
symmetries~\cite{Wilson}.  Because supersymmetry can not be broken
with this mechanism, only non-supersymmetric models are considered.

First, let us consider the 5-dimensional space-time $M^4\times S^1$
with coordinates $x^{\mu}$, and $y\equiv x^5$. The radius for the
fifth dimension is $R$. The gauge fields are denoted as $A_M (x^{\mu},
y)$ where $M=0, 1, 2, 3, 5$. Because $Z_2$ belongs to the fundamental
group of $S^1$, we can define a $Z_2$ parity operator $P_y$ for a
generic bulk multiplet $\Phi(x^{\mu}, y)$
\begin{eqnarray}
\Phi(x^{\mu},y)&\to \Phi(x^{\mu},y+ 2\pi R)=
\eta_{\Phi} P_y^{l_{\Phi}}\Phi(x^{\mu},y)
(P_y^{-1})^{m_{\Phi}}~,~\,
\end{eqnarray}
where $\eta_{\Phi} = \pm 1$ and $P_y^2=1$.

For a $SU(5)$ model, if we choose $P_y$ to be equal to $P'$ in
Eq.~(\ref{eq:pppsu5}), the $SU(5)$ gauge symmetry is broken down to the
SM gauge symmetry for the zero modes.

For a $SO(10)$ model, if we choose $P_y$ the same as
$P'$ in Eqs. (\ref{eq:psu5}), (\ref{eq:pfsu5}) or (\ref{eq:pps}), the
$SO(10)$ gauge symmetry is broken down to the $SU(5)\times U(1)$,
flipped $SU(5)\times U(1)'$, or the PS gauge symmetry respectively,
for the zero modes.

Similarly, we can consider the 6-dimensional space-time $M^4\times
S^1\times S^1$, with coordinates $x^{\mu}$, $y\equiv x^5$, and
$z\equiv x^6$. The radii for the fifth and sixth dimensions are $R_1$
and $R_2$ respectively.  We can define two $Z_2$ parity operators
$P_y$ and $P_z$ for a generic bulk multiplet $\Phi(x^{\mu}, y, z)$
\begin{eqnarray}
\Phi(x^{\mu},y, z)&\to \Phi(x^{\mu},y+ 2\pi R_1, z)=
\eta_{\Phi}^y P_y^{l_{\Phi}}\Phi(x^{\mu},y, z)
(P_y^{-1})^{m_{\Phi}}~,~\,
\end{eqnarray}
\begin{eqnarray}
\Phi(x^{\mu},y, z)&\to \Phi(x^{\mu},y, z+ 2\pi R_2)=
\eta_{\Phi}^z P_z^{l_{\Phi}}\Phi(x^{\mu},y, z)
(P_z^{-1})^{m_{\Phi}}~,~\,
\end{eqnarray}
where $\eta_{\Phi}^y=\pm 1$, $\eta_{\Phi}^z=\pm 1$,
$P_y^2=1$, and $P_z^2=1$.

By choosing the following $P_y$ and $P_z$ exactly the same as $P$ and
$P'$ in Eqs. (\ref{eq:psmx1}), (\ref{eq:psmx2}) or (\ref{eq:psmx3}),
we can break the $SO(10)$ down to the SM like gauge symmetry.

Inspired by the high dimensional non-supersymmetric GUTs with Wilson
line gauge symmetry breaking, we can construct the 4-dimensional $N=1$
supersymmetric $G^N$ models where the gauge symmetry $G^N$ can be
broken down to a diagonal subgroup of $G$ or the SM like gauge
symmetry. SUSY GUT models with Wilson line gauge symmetry breaking
have the problem that the 5-dimensional $N=1$ supersymmetry or the
6-dimensional $N=2$ supersymmetry can not be broken down to the
4-dimensional $N=1$ supersymmetry.  However, we can start from the
4-dimensional $N=1$ supersymmetry, which is exactly the advantage of
the 4-dimensional supersymmetric GUT models inspired by the
deconstructions.

\subsection{Deconstruction of Orbifold GUTs and Wilson Line Gauge
  symmetry Breaking} 

For the deconstruction of the orbifold SUSY GUTs, we break the
$G^{N-1}\times G_s$ to a diagonal $G_s \subset G$ as the following.
Suppose there are $N$ nodes, and the gauge symmetry on the first
$(N-1)$ nodes is $G$ while the gauge symmetry on the last node is
$G_s$.  The nodes are connected with $(N-1)$ bi-fundamental fields
$U_i$, and there is no link field between the first and last nodes.
When the $U_i$'s acquire uniform VEVs, $<U_i> = (v / {\sqrt{2}})~{\rm
  diag}(1,1,\cdots,1)$, for all $i$'s.  The gauge bosons of the group
$G_s$ have a $N \times N$ mass matrix, which has a zero eigenvalue,
while the $G/G_s$ gauge bosons have a $(N-1) \times (N-1)$ mass
matrix, which does not have a zero eigenvalue~\cite{Csaki:2001qm}.  As
the $G/G_s$ gauge bosons become heavy, $G^{N-1}\times G_s$ is
effectively broken down to $G_s$ for the massless fields.  Additional
fields will be needed on the first and $N$-th nodes to cancel
anomalies.  Alternatively, $N$ bi-fundamental fields $U_i$ connect the
$N$ nodes to form a loop.  With the same uniform VEVs, the $G_s$ gauge
bosons have $N \times N$ mass matrices, while the $G/G_s$ gauge bosons
have $(N-1) \times (N-1)$ mass matrices. Only one set of $G_s$ gauge
bosons remain massless.  The anomalies are always canceled in this
setup, thus no additional field is required.

From now on we always discuss the simplified case of two copies of the
initial $G$ group $G_1 \times G_2$ where $G_2 \subseteq G_1$, with
$A_\mu^\alpha$ and $T_\alpha$ being the gauge fields and generators of
$G_1$, and $B_\mu^\beta$ and $T_\beta$ the gauge fields and generators
of $G_2$. We assume that the gauge couplings are equal at the breaking
scale for simplicity.  The generalization to more copies of the $G$
group is straightforward.  With $U_1$ and $U_2$ being two
bi-fundamental fields, the covariant derivatives are
\bea
\label{eq:covderi}
D_\mu U_1 = \partial_\mu U_1 - i A_\mu^{\alpha} T_\alpha U_1 + i
B_\mu^{\beta} T_\beta U_1\,,\nn\\
D_\mu U_2 = \partial_\mu U_2 + i A_\mu^{\alpha} T_\alpha U_2 - i
B_\mu^{\beta} T_\beta U_2\,,
\eea
and the effective action for the scalar fields is
\bea
\label{eq:action}
 S&=&\int d^4x \left\{-{1\over {4 g^2}} Tr F_1^2 +
Tr[(D_{\mu} U_1)^{\dagger} D^{\mu} U_1] - {1\over {4 g^2}} Tr F_2^2 +
\right.\nonumber\\ && \left.
Tr[(D_{\mu} U_2)^{\dagger} D^{\mu} U_2]+...
\right\}~.~\,
\eea
For simplicity, we only write down the effective action for the scalar
fields.

Inspired by the high dimensional non-supersymmetric GUTs with Wilson
line gauge symmetry breaking, we can use the VEVs of the
bi-fundamental fields that are not commutating with a subset of the
generators to break the original group~\cite{Li:2002xd}.  Let us
suppose the initial gauge group is $G\times G$ and $G_s$ is a subgroup
of $G$.  The term $Tr[(D_{\mu} U_1)^{\dagger} D^{\mu} U_1] +
Tr[(D_{\mu} U_2)^{\dagger} D^{\mu} U_2]$ in Eq.~(\ref{eq:action})
produces a mass matrix of the form
\be
\label{eq:gmmatrx}
(\begin{array}{cc}
A_\mu^\alpha&B_\mu^\beta
\end{array})
\left(
\begin{array}{cc}
Tr[U_1^{\dagger}T_\alpha^{\dagger} T_{\alpha'} U_1+T_\alpha^{\dagger} U_2^{\dagger}
U_2 T_{\alpha'}] & -Tr[U_1^{\dagger} T_\alpha^{\dagger} U_1 T_{\beta'} +
T_\alpha^{\dagger} U_2^{\dagger} T_{\beta'} U_2] \\
-Tr[T_\beta^{\dagger} U_1^{\dagger} T_{\alpha'} U_1 + U_2^{\dagger}
T_\beta^{\dagger} U_2 T_{\alpha'}] & Tr[T_\beta^{\dagger} U_1^{\dagger} U_1
T_{\beta'} + U_2^{\dagger} T_\beta^{\dagger} T_{\beta'} U_2]
\end{array}
\right)
\left(
\begin{array}{c}
A_\mu^{\alpha'}\\
B_\mu^{\beta'}
\end{array}
\right) \, .
\ee

After $U_1$ and $U_2$ acquire VEVs, the $U_i$'s in the mass matrix
will be replaced by $<U_i>$.  The VEV $<U_i>$ has the property that
$<U_i>^{\dagger}<U_i>$ is always proportional to the identity matrix.
The normalized generators satisfy the relation $Tr(T_\alpha
T_{\alpha'}) = (1/2)~\delta_{\alpha\alpha'}$ for $SU(N)$ or
$Tr(T_\alpha T_{\alpha'}) = 2 \delta_{\alpha\alpha'}$ for $SO(N)$,
thus the diagonal terms in matrix in Eq.~(\ref{eq:gmmatrx}) are always
proportional to $\delta_{\alpha\alpha'}$, while the off-diagonal terms
depend on the commutation relations between the VEVs of the
bi-fundamental fields and the generators.  If the VEV of $U_1$
commutes with all the generators of $G$, while the VEV of the $U_2$
commutes with the generators of $G_s$ group, $T_a$, and anti-commutes
with those of $G/G_s$, $T_{\hat a}$
\be
<U_1>^{\dagger} T_a <U_1> = T_a \quad {\rm and} \quad <U_1>^{\dagger}
T_{\hat a}
<U_1> =  T_{\hat a} \, ,
\ee
\be
<U_2>^{\dagger} T_a <U_2> = T_a \quad {\rm and} \quad <U_2>^{\dagger}
T_{\hat a}
<U_2> = - T_{\hat a} \, ,
\ee
the mass matrices become proportional to
\be
(\begin{array}{cc}
A_\mu^a & B_\mu^b
\end{array})
\left(
\begin{array}{cc}
1 & -1 \\
-1 & 1
\end{array}
\right)
\left(
\begin{array}{c}
A_\mu^a \\
B_\mu^b
\end{array}
\right)
\quad {\rm and} \quad
(\begin{array}{cc}
{ A}_\mu^{\hat a} & { B}_\mu^{\hat b}
\end{array})
\left(
\begin{array}{cc}
1 & 0 \\
0 & 1
\end{array}
\right)
\left(
\begin{array}{c}
{ A}_\mu^{\hat a} \\
{ B}_\mu^{\hat b}
\end{array}
\right) \,, \ee respectively.  While the former matrix has a $0$
eigenvalue, the latter does not.  Thus the gauge symmetry $G\times G$
is broken down to the diagonal subgroup $G_s$.

Before presenting our models, we would like to emphasize that we
consider the 4-dimensional $N=1$ supersymmetry in all the models
discussed in this paper.

\section{Breaking via Pati-Salam}
\label{sec:patsal}

One way to realize the deconstruction of orbifold supersymmetric
$SO(10)$ models is to start with the initial gauge group $SO(10)\times
SO(6)\times SO(4)$. It is broken down to the diagonal PS gauge
symmetry by bifundamental link fields. The doublet-triplet splitting
can be achieved by the missing partner mechanism, and the proton decay
problem is solved because the gauge coupling unification scale is a
few times $10^{16}$ GeV. A simple variation of this model is the
$SO(10)\times SO(6)$ model.  In addition, we can consider the
$SO(10)\times SO(10)$ model, where the gauge symmetry can be broken
down to the diagonal PS gauge symmetry, which is inspired by the
deconstruction of 5-dimensional Wilson line gauge symmetry breaking.
However, the doublet-triplet splitting problem can not be solved
without fine-tuning in superpotential.

\subsection{$SO(10) \times SO(6) \times SO(4)$ Model}
\label{sec2.1}
To break $SO(10) \times SO(6) \times SO(4)$ to the diagonal PS gauge
symmetry, we rely on two isomorphisms, $SO(6) \cong SU(4)$ and $SO(4)
\cong SU(2) \times SU(2)$.  We introduce the bi-fundamental fields
$U_1$ and $U_2$ with the following quantum numbers
\begin{equation}
\label{so10xps}
\begin{array}{c|cc}
      & SO(10) &  SO(6) \times SO(4)\\
\hline
{\vrule height 15pt depth 5pt width 0pt}
 U_1       & {\bf 10} & ({\bf {6}}, {\bf 1})\\
 U_2       & {\bf 10} & ({\bf 1}, {\bf {4}})\\
\end{array} \nonumber
\end{equation}
Suppose, for example, we have the superpotential
\begin{eqnarray}
W  &=& S_{1} \left( U_1 U_1 - 3 v^2 \right)
+ S_{2} \left( U_2 U_2 - 2 v^2 \right)~,~\,
\end{eqnarray}
where $S_1$ and $S_2$ are singlets, and the link
fields acquire VEVs
\be
 <U_1> =  {v\over {\sqrt 2}} \left(
  \begin{array}{c}
    I_{6\times6} \\
    0_{4\times6} \\
  \end{array}
  \right)~, \quad
 <U_2>  =  {v\over {\sqrt 2}} \left(
  \begin{array}{c}
    0_{6\times4} \\
    I_{4\times4} \\
  \end{array}
  \right)~,
\ee
where $I_{i\times i}$ is the $i\times i$ identity matrix, and
$0_{i\times j}$ is a $i\times j$ matrix where all the entries are
zero. The $2 \times 2$ mass matrix for the PS gauge bosons is
\begin{equation}
\label{eq:sm32}
m_{PS}^2 = 2 g^2 v^2 X_1 \,,\quad X_1 \equiv
\left(
\begin{array}{cc}
1&-1\\
-1&1\\
\end{array} \right),
\end{equation}
with eigenvalues $0, 4 g^2v^2$.  The masses of the
$SO(10)/(SO(6)\times SO(4))$ gauge bosons are simply $ 2 g^2 v^2$.  If
generalized to a model with gauge group $SO(10)^{N-1}\times
SO(6)\times SO(4)$, the PS gauge bosons have a $N \times N$ mass
matrix which has a zero eigenvalue, whereas the non-PS gauge bosons
have a $(N-1) \times (N-1)$ mass matrix, for which the zero eigenvalue
is absent.

Three families of the SM fermions form three ${\bf 16}$ spinor
representations of $SO(10)$, and we introduce a 10-dimensional Higgs
field $H_{\bf 10}$.  To break the PS gauge symmetry and at the same
time require a left-right symmetry, \ie, the coupling $\alpha_L$ of
the $SU(2)_L$ equals to $\alpha_R$ of the $SU(2)_R$ above the PS gauge
symmetry unification scale ($M_{PS}$), we introduce the fields
$\Sigma_{h}$, ${\overline \Sigma}_{h}$, $\Sigma_{f}$, ${\overline
  \Sigma}_{f}$, $C_1$, $C_2$, $S$ and $S'$.  Their quantum numbers
under the gauge symmetry $SO(10) \times SU(4) \times SU(2)_L\times
SU(2)_R$ are given in the following
\begin{equation}
\label{so10epf}
\begin{array}{c|cc}
      & SO(10) &  SU(4) \times SU(2)_L\times SU(2)_R\\
\hline
{\vrule height 15pt depth 5pt width 0pt}
 \Sigma_{h}      & {\bf 1} & ({\bf {4}}, {\bf 1}, {\bf {2}})\\
 {\overline \Sigma}_{h}      & {\bf 1} & ({\bf {\bar 4}}, {\bf 1},
 {\bf {2}})\\
 \Sigma_{f}      & {\bf 1} & ({\bf {4}},  {\bf {2}}, {\bf 1})\\
 {\overline \Sigma}_{f}      & {\bf 1} & ({\bf {\bar 4}}, {\bf {2}},
 {\bf 1})\\
 C_1, C_2 & {\bf 1}  & ({\bf { 6}}, {\bf {1}}, {\bf 1})\\
 S, S' & {\bf 1}  & ({\bf { 1}}, {\bf {1}}, {\bf 1})\\
\end{array} \nonumber
\end{equation}
Note that $\Sigma_{h}$, ${\overline \Sigma}_{h}$, $\Sigma_{f}$,
${\overline \Sigma}_{f}$ can form one pair of ${\bf 16}$ and ${\bf
  \overline {16}}$ under $SO(10)$. The PS gauge symmetry is broken
down to the SM gauge symmetry when the right-handed neutrino scalar
component in $\Sigma_{h}$ and its charge conjugation in ${\overline
  \Sigma}_{h}$ obtain VEVs at the $M_{PS}$ scale.

The superpotential  is
\begin{eqnarray}
W  = && y_1 S (\Sigma_{h} {\overline \Sigma}_{h} -M_{PS}^2)
+M_{PS} \Sigma_{f} {\overline \Sigma}_{f}
+y_2 \Sigma_{h} C_2 \Sigma_{h} + \nn \\
&& y_3 {\overline \Sigma}_{h} C_2 {\overline \Sigma}_{h}
+ y_4 H_{\bf 10} U_1 C_1 + y_5 S' H_{\bf 10} H_{\bf 10}
+ y_{ij} 16_i H_{\bf 10} 16_j~,~\,
\label{superpotential}
\end{eqnarray}
where $y_i$'s are Yukawa couplings, and $y_{ij}$'s are the usual
Yukawa couplings for the SM fermions.  Moreover, similarly to the
doublet-triplet splitting via the missing partner mechanism, the color
triplets in $H_{\bf 10}$ and $C_1$ obtain the GUT scale masses while
the doublets are massless because of the superpotential term $y_4
H_{\bf 10} U_1 C_1$. Subsequently, the proton decay due to the
dimension-5 operator is negligible because the mixing of the color
triplets through the $\mu$ term is very small.  Limits on the
contributions from the dimension-6 proton decay operators require that
the gauge coupling unification scale be larger than $5\times 10^{15}$
GeV~\cite{Giudice:2004tc}, which is obviously satisfied in our model
where the GUT scale is at least a few times $10^{16}$ GeV.
Furthermore, the $\mu$ term for the Higgs doublets can be generated
from the superpotential $y_5 S' H_{\bf 10} H_{\bf 10}$, as is similar
to the next to the minimal supersymmetric standard model~\cite{NMSSM}.
We also assume that the masses for $C_2$, $\Sigma_{h}$ and ${\overline
  \Sigma}_{h}$ are around $M_{PS}$, and the masses for $C_1$ and Higgs
triplets in $H_{\bf 10}$ are about ${\sqrt 2} g v$.  We do not discuss
the fermion masses and mixings, as they are out of the scope of this
paper.

For simplicity, we assume the universal masses $M_{SUSY} = 500$ GeV
for the supersymmetric particles~\footnote{As is well-known, the
  masses at the low scale for supersymmetric particles can be
  determined from SUSY breaking soft terms at the high scale by
  renormalization group equation running. For our purpose in this
  paper, it is not necessary to carry out this running.}. From $M_{Z}$
to $M_{SUSY}$, the gauge couplings evolve the same way as those in the
SM, and the beta functions are $b^0 \equiv (b_1, b_2, b_3) =
(41/10,-19/6,-7)$, where $b_1$, $b_2$ and $b_3$ are for the gauge
symmetries $U(1)_Y$, $SU(2)_L$ and $SU(3)_C$, respectively. From
$M_{SUSY}$ to the PS gauge symmetry unification scale $M_{PS}$, the
beta functions are just those of the MSSM, {\it i.e.}, $b^I =
(33/5,1,-3)$.

At the PS gauge symmetry
unification scale $M_{PS}$, the gauge couplings for $U(1)_Y$ and $SU(3)_C$ are
 \be \alpha_1^{-1}(M_{PS}) =
\alpha_1^{-1}(M_Z) - \frac{b_1^0}{2\pi}
\log\left(\frac{M_{SUSY}}{M_Z}\right)
-\frac{b_1^I}{2\pi}
\log\left(\frac{M_{PS}}{M_{SUSY}}\right)~,
\ee
\be
\alpha_3^{-1}(M_{PS}) = \alpha_3^{-1}(M_Z) - \frac{b_3^0}{2\pi}
\log\left(\frac{M_{SUSY}}{M_Z}\right)
-\frac{b_3^I}{2\pi}
\log\left(\frac{M_{PS}}{M_{SUSY}}\right)~.
\ee
The $SU(2)_R$ gauge coupling
$\alpha_R$ is related to $\alpha_1$ and $\alpha_3$ by
\be
\alpha_R^{-1}(M_{PS}) = \frac{5}{3} \alpha_1^{-1}(M_{PS}) -
\frac{2}{3} \alpha_3^{-1}(M_{PS})~.
\ee

Above the PS scale, including the contributions from fermions, PS
gauge multiplets, two Higgs doublets $H^D$, $\Sigma_{h}$ ${\overline
  \Sigma}_{h}$, $\Sigma_f$, ${\overline \Sigma}_f$, and $C_2$, the
beta functions $b^{II} \equiv (b_R,b_L,b_4)$ become
\bea
b^{II} &=& b(M) + b^{PS}(V) + b(H^D) + b(\Sigma) + b(C_2) \nn \\
&=& (6,6,6) + (-6,-6,-12) + (1,1,0) + (4, 4, 4) + (0,0,1) = (5,5,-1)~,
\eea
where $b_R$, $b_L$ and $b_4$ are the beta functions for $SU(2)_R$,
$SU(2)_L$ and $SU(4)_C$, respectively.  In the region above $\sqrt{2}
gv$, one set of $SO(10)/PS$ gauge multiplets, $C_1$ and
Higgs triplets $H^T$ in $H_{\bf 10}$ appear, then the beta functions are
\bea
b^{III} &=& b^{II} + b^{SO(10)/PS}(V) + b(C_1) + b(H^T) \nn \\
  &=& (5,5,-1) + (-18,-18,-12) + (0, 0, 1) + (0, 0, 1)
  = (-13,-13,-11).
\eea

\begin{figure}[ht]
\centerline{\includegraphics[width=8cm]{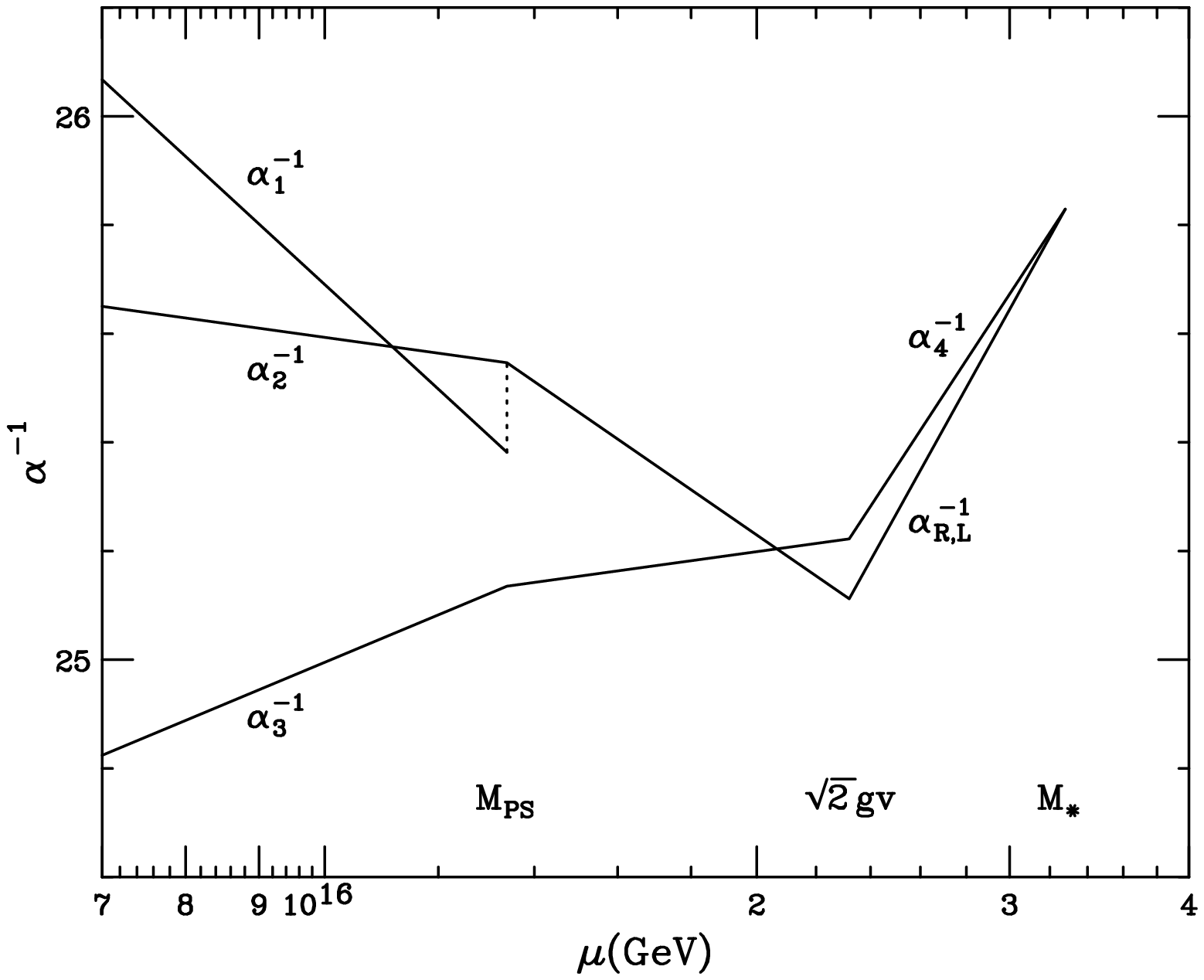}
            \includegraphics[width=8cm]{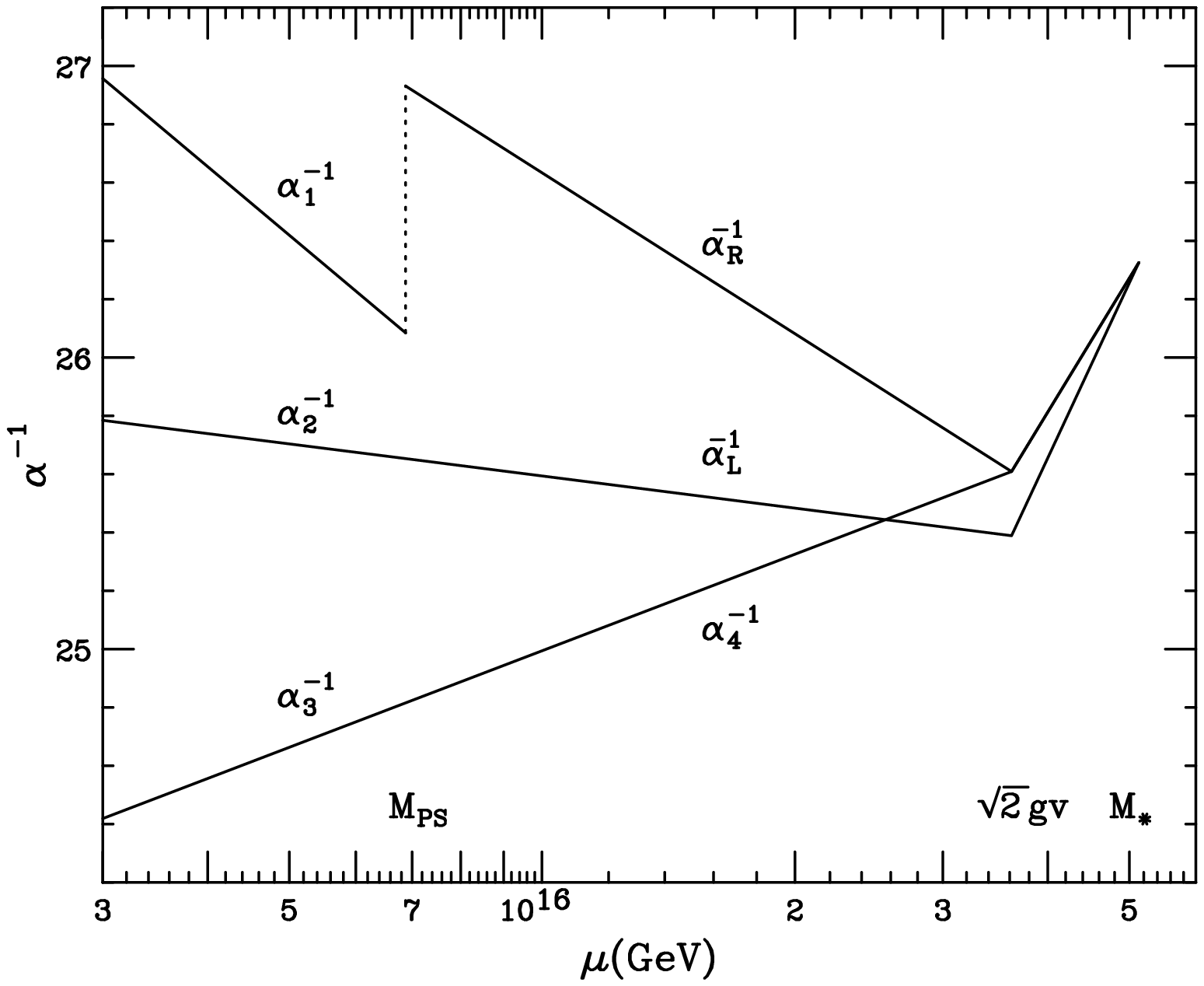}}
\caption[]{Left: The gauge coupling unification for the left-right
  symmetric PS model near $M_* = 3.3 \times 10^{16}$ GeV.  Right: The
  gauge coupling unification for the left-right non-symmetric PS model
  near $M_* = 5.1 \times 10^{16}$ GeV.
\label{fig:uni_ps1}}
\end{figure}

We show the runnings of the gauge couplings near the unification scale
in the left panel of Fig.~\ref{fig:uni_ps1}.  Note that the runnings
of $\alpha_L$ and $\alpha_R$ coincide above $M_{PS}$ because we have
required the left-right symmetry.  The gauge coupling $\alpha_4$ of
$SU(4)$ unify with $\alpha_R$ and $\alpha_L$ at the scale $M_* = 3.3
\times 10^{16}$ GeV and the PS scale occurs at $M_{PS} = 1.3 \times
10^{16}$ GeV. Thus, there is no proton decay problem.
In the above construction, we maintain the left-right symmetry above
the $M_{PS}$ scale.  If we no longer require the left-right symmetry,
we can drop the additional fields $\Sigma_f$ and ${\overline
  \Sigma}_f$ from the table in Eq.~(\ref{so10epf}).  In this setup,
the discussion of the doublet-triplet splitting is unchanged from that
of the above symmetric case.  The RGE runnings of the gauge couplings
are different due to the simpler field content.  More specifically,
the beta functions are, $b^{II} = (5,1, -3)$ and $b^{III} =
(-13,-17,-13)$.  Above $M_{PS}=6.9\times 10^{15}$ GeV, $\alpha_R$ and
$\alpha_L$ have different RGE runnings, as is evident in
Fig.~\ref{fig:uni_ps1} right panel.  With a GUT scale of $M_* = 5.1
\times 10^{16}$ GeV, the proton lifetime is safely above the current
experimental bound.

\subsection{$SO(10)\times SO(6)$ Model}

If we want to solve the proton decay problem, we can also consider the
model with gauge group $SO(10)\times SO(6)$, which can not be obtained
but is inspired by the orbifold GUT.

To break $SO(10) \times SO(6)$ to PS gauge symmetry,
 we introduce bi-fundamental field
$U_1$ as
\begin{equation}
\label{so10xso6}
\begin{array}{c|cc}
      & SO(10) &  SO(6) \\
\hline
{\vrule height 15pt depth 5pt width 0pt}
 U_1       & {\bf 10} & ({\bf {6}}, {\bf 1})\\
\end{array} \nonumber
\end{equation}
The link
fields acquire the following VEVs
\be
 <U_1> =  {v\over {\sqrt 2}} \left(
  \begin{array}{c}
    I_{6\times6} \\
    0_{4\times6} \\
  \end{array}
\right)~.  
\ee 
The $SO(10)\times SO(6)$ gauge symmetry is broken down
to the PS gauge symmetry.

The discussions on the proton decay, PS gauge symmetry breaking and
the gauge coupling unification are similar to those in the above
subsection, so, we will not repeat them here. However, we would like
to emphasize that the initial gauge group is smaller than that in the
above subsection.

\subsection{$SO(10)\times SO(10)$ Model}

Inspired by the deconstruction of 5-dimensional $SO(10) $ model with
Wilson line gauge symmetry breaking, we consider the model with the
simplest gauge group $SO(10)\times SO(10)$.  Let us introduce
bi-vector fields $U_1$ and $U_2$ with the following quantum numbers
\begin{equation}
\label{so10xso10}
\begin{array}{c|cc}
      & SO(10) & SO(10) \\
\hline
{\vrule height 15pt depth 5pt width 0pt}
 U_1       & {\bf 10} & {\bf 10}\\
 U_2       & {\bf 10} & {\bf 10}\\
\end{array} \nonumber
\end{equation}

The effective action is the same as that in Eq.~(\ref{eq:action}).  If
the $U_1$ and $U_2$ fields acquire VEVs
\be
 <U_1> = {v\over \sqrt{2}}  \,\sigma_0 \otimes {\rm diag} ( 1, 1,
 1,1,1)\,,\quad
 <U_2> = {v\over \sqrt{2}} \,\sigma_0 \otimes {\rm diag}
 (-1,-1,-1,1,1)\,,
\ee
the $2 \times 2$ mass matrix for the PS gauge bosons is
\begin{equation}
\label{eq:so102}
m_{PS}^2 = 4 g^2 v^2 X_1,
\end{equation}
with $X_1$ defined in Eq.~(\ref{eq:sm32}).  The squared masses for the
PS group gauge bosons are either $0$ or $8 g^2v^2$.  The mass matrix
of the non-PS ($SO(10)/PS$) gauge bosons is
\begin{equation}
\label{eq:so2nops}
m_{NPS}^2 = 4 g^2 v^2 X_2\,, \quad X_2 \equiv
\left(
\begin{array}{cc}
1&0\\
0&1\\
\end{array} \right).
\end{equation}
It has degenerate eigenstates with squared masses equal to $4 g^2v^2$.

Three families of the SM fermions form three ${\bf 16}$ spinor
representations under the first $SO(10)$ while they are singlets under
the second $SO(10)$.  We also introduce a Higgs field $H_{\bf 10}$. To
give masses to the color triplets in the Higgs field $H_{\bf 10}$, we
introduce a 10-dimensional field $H'_{\bf 10}$, and to break the PS
gauge symmetry, we introduce the fields $\Sigma_{H}$, ${\overline
  \Sigma}_{H}$, and two singlets $S$ and $S'$.  The quantum numbers
for these particles are
\begin{equation}
\label{so10epf22}
\begin{array}{c|cc}
      & SO(10) &  SO(10)\\
\hline
{\vrule height 15pt depth 5pt width 0pt}
{\bf 16}_i & {\bf 16} & {\bf 1} \\
H_{\bf 10} & {\bf 10} & {\bf 1} \\
H'_{\bf 10} & {\bf 1} & {\bf 10} \\
 \Sigma_{H}      & {\bf 16} & {\bf 1} \\
 {\overline \Sigma}_{H}  & {\bf {\overline 16}} & {\bf 1}    \\
S, S' & {\bf 1}  & {\bf { 1}}\\
\end{array} \nonumber
\end{equation}

The superpotential  is
\begin{eqnarray}
W  = && y_1 S (\Sigma_{H} {\overline \Sigma}_{H} -M_{PS}^2)
+ y_2 H_{\bf 10} (U_1 -U_2) H'_{\bf 10} \nn \\
&& +M_* H'_{\bf 10} H'_{\bf 10}
+ y_3 S' H_{\bf 10} H_{\bf 10}
+ y_{ij} 16_i H_{\bf 10} 16_j~,~\,
\label{superpotential22}
\end{eqnarray}
where $M_* $ is the $SO(10)$ unification scale.  The doublet-triplet
splitting can be obtained through the fine-tunning superpotential $y_2
H_{\bf 10} (U_1 -U_2) H'_{\bf 10}$, as in the usual $SU(5)$ model.
Because it is less interesting than the usual $SO(10)$ model from the
phenomenological point of view, we do not study it in detail.

\section{Breaking via $SU(5) \times
  U(1)$ or  Flipped $SU(5) \times U(1)'$}
\label{sec:noncom}
For the $SO(10)\times {\rm flipped ~} SU(5) \times U(1)'$ and $SO(10)
\times SU(5) \times U(1)$ models from the deconstruction of
5-dimensional supersymmetric orbifold $SO(10)$ models, the discussions
on how to break the gauge symmetry down to the diagonal flipped $SU(5)
\times U(1)'$ or $SU(5) \times U(1)$ gauge symmetry are identical to
those of Section~\ref{sec2.1}.  Two link fields with simple diagonal
VEVs will leave the diagonal subgroup gauge bosons massless and give
the other gauge bosons masses of order $g v$.  We will consider the
models inspired by deconstruction of the 5-dimensional $SO(10)$ models
with Wilson line gauge symmetry breaking, where the $SO(10)$ gauge
symmetry can be broken down to the flipped $SU(5) \times U(1)'$ and
$SU(5) \times U(1)$, and the 6-dimensional $SO(10)$ model where the
$SO(10)$ gauge symmetry can be broken down to the SM like gauge
symmetry.

Starting with the $G \equiv SO(10) \times SO(10)$, we use two
bi-vectors $U_1$ and
$U_2$ with quantum numbers $({\bf 10}, {\bf 10})$.

In order to break the gauge symmetry down to the diagonal
$\su5u1$, the VEVs of the link fields are assumed to be
\bea
 <U_1> &=& {v\over \sqrt{2}} \sigma_0 \otimes {\rm diag} (1,1,1,1,1) ~,~\nn\\
 <U_2> &=& {v\over \sqrt{2}} \sigma_2 \otimes {\rm diag} (1,1,1,1,1) ~,~\nn
\eea and to obtain the flipped $\fsu5u1$, $U_2$ VEV is replaced
with
\be
 <U_2> = {v\over \sqrt{2}} \sigma_2 \otimes {\rm diag} (1,1,1,-1,-1) ~.~
\ee

Note the similarity between the VEVs and the parity operators in Eqs.
(\ref{eq:psu5}) and (\ref{eq:pfsu5}).  As $<U_1>$ commutes with $T_a$
and $T_{\hat a}$, while $<U_2>$ commutes with $T_a$ and anti-commutes
with $T_{\hat a}$, the gauge bosons of the unbroken subgroup $G_s =
\su5u1$ or flipped $\fsu5u1$ have a mass matrix identical to the right
hand side of Eq.~(\ref{eq:so102}) while the $G/G_s$ gauge bosons have
a mass matrix the same as the right hand side of
Eq.~(\ref{eq:so2nops}).  One set of the $G_s$ gauge bosons remain
massless and the $G/G_s$ gauge bosons acquire masses of $2gv$.

Moreover, the $\so10\times \so10$ can be broken to the intersection of two maximal
subgroups, $SU(3)_C \times SU(2)_L \times U(1)_Y \times U(1)_X$.  If we
introduce four bi-vector link fields with VEVs
\bea
<U_1> &=& {v\over \sqrt{2}} \sigma_0 \otimes {\rm diag} (1,1,1,1,1) ~,~\nn\\
<U_2> &=& {v\over \sqrt{2}} \sigma_2 \otimes {\rm diag} (1,1,1,1,1) ~,~\nn\\
<U_3> &=& {v\over \sqrt{2}} \sigma_0 \otimes {\rm diag} (1,1,1,1,1) ~,~\nn\\
<U_4> &=& {v\over \sqrt{2}} \sigma_0 \otimes {\rm diag} (1,1,1,-1,-1)
~,~ 
\eea 
where all $U_i$'s are $({\bf 10},{\bf 10})$ under the
original group $SO(10) \times SO(10)$.  It is easy to see that $U_1$
and $U_2$ are the same as the link fields we used for breaking $SO(10)
\times SO(10)$ to $\su5u1$, and $U_3$ and $U_4$ are the same as those
for breaking $SO(10) \times SO(10)$ to the PS group.  To clarify the
notation, we assign $m_{SMX}$ for the mass matrices of the gauge
bosons in the intersection of the $SU(5)\times U(1)$ and PS gauge
symmetry, which is $SU(3)_C\times SU(2)_L \times U(1)_Y \times
U(1)_X$, $m_5$ for those in $\su5u1$ but not in PS gauge group,
$m_{PS}$ for those in the PS gauge group but not in $\su5u1$, and
$m_R$ is reserved for the mass matrices of the rest gauge bosons
belong to neither $SU(5)\times U(1)$ or PS gauge group. These mass
matrices are given by \be m_{SMX}^2 = 4 g^2 v^2 \left(
\begin{array}{cc}
2&-2\\
-2&2\\
\end{array} \right) ~,
\quad
m_{5}^2 = m_{PS}^2 = 4 g^2 v^2 \left(\begin{array}{cc}
2&-1\\
-1&2\\
\end{array} \right) ~,
\quad
m_{R}^2 = 4 g^2 v^2 \left(
\begin{array}{cc}
2&0\\
0&2\\
\end{array} \right) ~.
\ee

It is obvious that only $m_{SMX}$ has 0 eigenvalue, and the others do
not.  With the combination of these two sets of link fields, the
$\so10 \times \so10$ gauge symmetry is broken to the $SU(3)_C\times
SU(2)_L \times U(1)_Y \times U(1)_X$ gauge symmetry.

Similarly, we can break $\so10 \times \so10$ down
to $SU(3)_C\times SU(2)_L \times U(1)_Y \times U(1)_X$ by choosing
\bea
<U_1> &=& {v\over \sqrt{2}} \sigma_0 \otimes {\rm diag} (1,1,1,1,1) ~,~\nn\\
<U_2> &=& {v\over \sqrt{2}} \sigma_2 \otimes {\rm diag} (1,1,1,-1,-1) ~,~\nn\\
<U_3> &=& {v\over \sqrt{2}} \sigma_0 \otimes {\rm diag} (1,1,1,1,1) ~,~\nn\\
<U_4> &=& {v\over \sqrt{2}} \sigma_0 \otimes {\rm diag} (1,1,1,-1,-1) ~,~
\eea
or
\bea
 <U_1> &=& {v\over \sqrt{2}} \sigma_0 \otimes {\rm diag} (1,1,1,1,1) ~,~\nn\\
 <U_2> &=& {v\over \sqrt{2}} \sigma_2 \otimes {\rm diag} (1,1,1,1,1) ~,~\nn\\
 <U_3> &=& {v\over \sqrt{2}} \sigma_0 \otimes {\rm diag} (1,1,1,1,1) ~,~\nn\\
 <U_4> &=& {v\over \sqrt{2}} \sigma_2 \otimes {\rm diag} (1,1,1,-1,-1) ~.~\,
\eea

In these models, the doublet-triplet splitting is still a problem
without fine-tuning in the superpotential, thus it is difficult to
solve the proton decay problem induced by the dimension-5 proton decay
operators.

\section{$SO(10)\times SO(10)$ Models with Bi-Spinor Link Fields}
\label{sec:bispinor}

In this section, we demonstrate that, by using bi-spinor link fields
(${\bf 16, {\overline{16}}}$) and (${\bf {\overline{16}}, 16 }$), we
are able to break the gauge symmetry $SO(10)\times SO(10)$ down to the
diagonal $SU(5)\times U(1)$ and flipped $\fsu5u1$ gauge symmetry, and
at the same time solve the doublet-triplet splitting problem via the
missing partner mechanism in the model with the flipped $\fsu5u1$
breaking chain.  In this approach, we do not rely on the commutation
relations between the link field VEVs and the generators.

We introduce two bi-fundamental fields $U_1$ and $U_2$
with following quantum numbers under $\sq20$
\begin{equation}
\label{eq:v1616_1}
\begin{array}{c|cc}      & SO(10) & SO(10) \\
\hline
{\vrule height 15pt depth 5pt width 0pt}
 U_1       & {\bf 16} & {\bf \overline{16}}\\
 U_2       & {\bf \overline{16}} & {\bf 16}\\
\end{array} \nonumber
\end{equation}

Then, the covariant derivative can be rewritten as, using $\sla$
defined in Appendix~\ref{apdx},
\bea
D_\mu U_1 = \partial_\mu U_1 -  \frac{i}{\sqrt{2}~} \sla U_1 +
\frac{i}{\sqrt{2}~} \slb U_1\,,\nn\\
D_\mu U_2 = \partial_\mu U_2 + \frac{i}{\sqrt{2}~} \sla U_2 -
\frac{i}{\sqrt{2}~} \slb U_2\,.
\eea

\subsection{$SU(5) \times U(1)$}
The effective action for scalar components is the same as that in
Eq.~(\ref{eq:action}), and the bi-spinors $U_1$ and $U_2$ acquire the
following VEVs
\be <U_1> = <U_2> = {v \over {\sqrt 2}}
\,{\rm diag}(0,0,0,1,0,0,0,1,1,1,1,0,0,0,0,1)\,.
\ee

Note that the assignment of the $1$'s are consistent with the diagonal
(${\bf {\bar 5}}, {\bf 5}$) and (${\bf 1}$, ${\bf 1}$) in the (${\bf
  16}, {\bf \overline{16}}$), see Apppendix~\ref{apdx}.  The gauge
fields in Eq.~(\ref{eq:gauge16}) have the mass matrices of the
following forms
\bea
m_{\lambda}^2 =
m_{V}^2= m_{W_L}^2= m_Y^2 = g^2 v^2 X_1 \,, \quad {\rm and} \quad
m_A^2 = m_{W_R}^2 = m_{X}^2 =2 g^2 v^2 X_2\,.
\eea

Thus, the massless fields are the 5 states in the $\lambda$ fields, 6
$V$, 2 $W_L$ and 12 $Y$ states, total of 25 independent states.  They
are the gauge bosons of the diagonal $\su5u1$ group.

\subsection{Flipped $\fsu5u1$}

For the flipped $\fsu5u1$, the ${\bf \bar{5}}$ and ${\bf 1}$ consist
of different fermion fields.  The VEVs of $U_1$ and $U_2$ are chosen
as
\be <U_1> = <U_2> = {v \over {\sqrt 2}} \,{\rm
  diag}(0,0,0,1,0,0,0,1,0,0,0,1,1,1,1,0)\,.  
\ee 
The mass matrices become \bea m_{\lambda}^2 = m_{V}^2= m_{W_L}^2=
m_A^2 = g^2 v^2 X_1 \,, \quad {\rm and} \quad m_Y^2 = m_{W_R}^2 =
m_{X}^2 =2 g^2 v^2 X_2\,, \eea and the massless states are 12 $A$, 2
$W_L$, 6 $V$ and the 5 states in $\lambda$'s.  These 25 massless
states are the gauge bosons of the diagonal flipped $\fsu5u1$.

Three families of the SM fermions form three ${\bf 16}$ spinor
representations under the first $SO(10)$ while they are singlets under
the second $SO(10)$.  We also introduce a 10-dimensional Higgs field
$H_{\bf 10}$, and one pair of Higgs $\Sigma_{H}$ and ${\overline
  \Sigma}_{H}$ in the spinor representation.  Moreover, we introduce
one pair of the fields $\chi$ and ${\overline \chi}$ in the spinor
representation, two singlet fields $S$ and $S'$.  The quantum numbers
for these particles are
\begin{equation}
\label{so10epf33}
\begin{array}{c|cc}
      & SO(10) &  SO(10)\\
\hline
{\vrule height 15pt depth 5pt width 0pt}
{\bf 16}_i & {\bf 16} & {\bf 1} \\
H_{\bf 10} & {\bf 10} & {\bf 1} \\
 \Sigma_{H}      & {\bf 16} & {\bf 1} \\
 {\overline \Sigma}_{H}  & {\bf {\overline {16}}} & {\bf 1}    \\
\chi & {\bf 1} & {\bf 16} \\
{\overline \chi} & {\bf 1} & {\bf {\overline {16}}} \\
S, S' & {\bf 1}  & {\bf { 1}}\\
\end{array} \nonumber
\end{equation}

The superpotential is
\begin{eqnarray}
W &=& y_1 \Sigma_{H} U_2 {\overline \chi}
+ y_2 {\overline \Sigma}_{H} U_1 \chi
+y_3 S (\Sigma_{H} {\overline \Sigma}_{H} -M_{\rm FS}^2)
+y_4 S' H_{\bf 10} H_{\bf 10}
+ \lambda_x \Sigma_{H} H_{\bf 10} \Sigma_{H}
\nonumber\\ &&
+\lambda_y {\overline \Sigma}_{H} H_{\bf 10} {\overline \Sigma}_{H}
+ y_{ij} 16_i H_{\bf 10} 16_j
+ {{y_5}\over {M_{\rm Pl}}} (U_1 U_2) ({\overline \chi} \chi)
~,~\,
\label{superpotential33}
\end{eqnarray}
where $M_{\rm FS} $ is the unification scale for the flipped
$\fsu5u1$, and $M_{\rm Pl}$ is the Planck scale.  The (${\bf \bar 5}
\oplus {\bf 1}$) multiplets in $\Sigma_{H}$ and $\chi$, and the (${\bf
  5} \oplus {\bf 1}$) multiplets in ${\overline \Sigma}_{H}$ and
${\overline \chi}$ are assumed to have the $SO(10)$ unification scale
masses from the first two terms in the superpotential, while the ${\bf
  10}$ multiplet $\Sigma_{h}$ in $\Sigma_{H}$, the ${\bf 10}$
multiplet ${\chi}_T$ in $\chi$, the ${\bf {\overline {10}}}$ multiplet
${\overline \Sigma}_{h}$ in ${\overline \Sigma}_{H}$, and the ${\bf
  {\overline {10}}}$ multiplet ${\overline \chi}_T$ in ${\overline
  \chi}$ are still massless at $M_*$.  For simplicity, we assume that
the masses $M_{\chi_T}$ for ${\chi}_T$ and ${\overline \chi}_T$ are
about $10^{12}$ GeV from the last term in superpotential.  Under
flipped $\fsu5u1$, $H_{\bf 10}$ is decomposed into one pair of 5-plets
$h$ and $\bar h$.  To be explicit, we denote $\Sigma_{h}$, ${\overline
  \Sigma}_{h}$, $h$ and $\bar h$ as
\begin{eqnarray}
\Sigma_{h}=(Q_H, D_H^c, N_H),
~{\overline \Sigma}_{h}= ({\bar {Q}}_{\bar H}, {\bar {D}}_{\bar H}^c,
{\bar {N}}_{\bar H}),
\label{Higgse2}
\end{eqnarray}
\begin{eqnarray}
h=(D_h^r, D_h^g, D_h^b, H_d),~{\bar h}=({\bar {D}}_{\bar h}^r, {\bar
  {D}}_{\bar h}^g, {\bar {D}}_{\bar h}^b, H_u).
\label{Higgse3}
\end{eqnarray}

From the third, fifth and sixth terms in Eq.~(\ref{superpotential33}),
we obtain the superpotential below $M_*$
\begin{eqnarray}
{\it W} = y_3 S ({\overline \Sigma}_{h} \Sigma_{h} -M_{\rm FS}^2) +
\lambda_x \Sigma_{h} \Sigma_{h} h +
 \lambda_y {\overline \Sigma}_{h}  {\overline \Sigma}_{h} {\bar h}~.
\label{spgut}
\end{eqnarray}
There is only one F-flat and D-flat direction, which can always be
rotated along the $N_H$ and ${\bar N}_{\bar H}$ directions. Hence we
obtain that $<N_H>=<{\bar N}_{\bar H}>=M_{\rm FS}$. In addition, the
superfields in $\Sigma_{h}$ and ${\overline \Sigma}_{h}$ are eaten or
acquire large masses via the supersymmetric Higgs mechanism, except
for $D_H^c$ and ${\bar D}_{\bar H}^c$. The superpotential $\lambda_x
\Sigma_{h} \Sigma_{h} h$ and $\lambda_y {\overline \Sigma}_{h}
{\overline \Sigma}_{h} {\bar h}$ couple the $D_H^c$ and ${\bar
  D}_{\bar H}^c$ with the $D_h$ and ${\bar {D}}_{\bar h}$ respectively
to form the massive eigenstates with masses $\lambda_x <N_H>$ and
$\lambda_y <{\bar N}_{\bar H}>$.  We naturally have the
doublet-triplet splitting due to the missing partner
mechanism~\cite{smbarr, dimitri, Huang:2003fv}.  Because the triplets
in $h$ and ${\bar h}$ only have small mixing through the $\mu$ term,
the higgsino-exchange mediated proton decay are negligible, {\it
  i.e.}, we do not have the dimension-5 proton decay problem. Proton
decay via the dimension-6 operators is well above the current
experimental bounds.

The gauge coupling unification for the flipped $SU(5) \times U(1)'$ is
realized by first unifying $\alpha_2$ and $\alpha_3$ at scale
$M_{23}$, then the gauge couplings of $SU(5)$ and $U(1)'$ further
unify at $M_*$.  From $M_{Z}$ to $M_{SUSY}$, the beta functions are
$b^0 \equiv (b_1, b_2, b_3) = (41/10,-19/6,-7)$, and from $M_{SUSY}$
to $M_{\chi_T} = 10^{12}$ GeV, the beta functions are $b^I =
(33/5,1,-3)$.  From $M_{\chi_T}$ to the $\alpha_2$ and $\alpha_3$
unification scale $M_{23}$, the beta functions are $b^{II} =
(36/5,4,0)$, because $\chi_T$ and ${\overline \chi}_T$ contribute to
the gauge coupling RGE runnings.

Unification
of $\alpha_2$ and $\alpha_3$ at $M_{23}$ gives the condition
\begin{eqnarray}
\alpha_2^{-1}(M_Z) - \alpha_3^{-1}(M_Z) = &&
 \frac{b_2^{0} - b_3^{0}}{2
  \pi} \log \left(\frac{M_{SUSY}}{M_Z}\right)
+ \frac{b_2^{I} - b_3^{I}}{2
  \pi} \log \left(\frac{M_{\chi_T}}{M_{SUSY}}\right)
\nonumber\\ &&
+\frac{b_2^{II} - b_3^{II}}{2
  \pi} \log \left(\frac{M_{23}}{M_{\chi_T}}\right)~,
\label{eq:su5uni1}
\end{eqnarray}
which can be solved for $M_{23}$.

The coupling $\alpha_1'$ of $U(1)'$ is related to $\alpha_1$ and
$\alpha_5$ at $M_{23}$ by
\be
\alpha_1'^{-1}(M_{23}) = \frac{25}{24} \alpha_1^{-1}(M_{23}) -
\frac{1}{24} \alpha_5^{-1}(M_{23})~.
\ee

Between $M_{23}$ and $M_*$, besides the 3 families of the Standard
Model fermions ${\bf 16}_i$, there are the $\fsu5u1$ gauge bosons in
${\bf 24} + {\bf 1}$, $H_{\bf 10}$, $\Sigma_h$,
${\overline{\Sigma}}_h$, $\chi_T$ and ${\overline \chi}_T$, thus the
beta functions for $U(1)'$ and $SU(5)$ are $b^{III} \equiv (b_1', b_5)
= (8, -2)$.

\begin{figure}[ht]
  \centerline{\includegraphics[width=8cm]{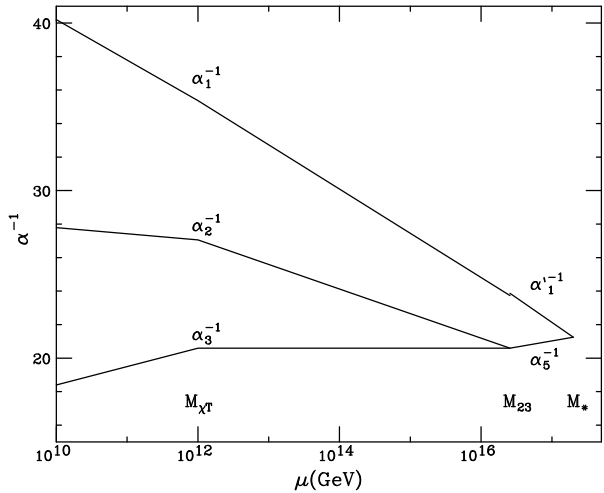}}
  \caption[]{The gauge coupling
unification near $M_* = 2.0 \times 10^{17}$ GeV for
    flipped $\fsu5u1$ model.
  \label{fig:flipuni}}
\end{figure}

In Fig.~\ref{fig:flipuni}, we show the runnings of the gauge couplings
near the unification scale $M_*$.  The unification scale $M_{23}$ for
$SU(2)_L$ and $SU(3)_C$ is about $2.6 \times 10^{16}$ GeV, and the
$SO(10)\times SO(10)$ unification scale $M_*$ is about $2.0 \times
10^{17}$ GeV.  If the masses $M_{\chi_T}$ for $\chi_T$ and ${\overline
  \chi}_T$ are larger than $M_{23}$, we will have $\alpha_1' >
\alpha_5$ at $M_{23}$. Because above $M_{23}$ the beta function of
$U(1)'$ is positive while that of $SU(5)$ is negative, we can not
achieve the unification of these two gauge couplings.  In short, the
intermediate mass scale $M_{\chi_T}$ is important for the gauge
coupling unification~\cite{Lopez:1993qn}.

\subsection{Brief Comments on $SO(10)\times {\rm Flipped} ~SU(5)\times U(1)'$}

We introduce the link fields $U_1$, $U_2$, $U_3$ and $U_4$ with
following quantum numbers
\begin{equation}
\label{so10FSU5}
\begin{array}{c|cc}
      & SO(10) & SU(5)\times U(1)' \\
\hline
{\vrule height 15pt depth 5pt width 0pt}
 U_1       & {\bf 16} & ({\bf {5}}, {\bf 3})\\
 U_2       & {\bf {\overline{16}}} & ({\bf {\bar 5}}, {\bf -3})\\
 U_3       & {\bf 16} & ({\bf 1}, {\bf {-5}})\\
 U_4     & {\bf {\overline{16}}} & ({\bf 1}, {\bf 5})\\
\end{array} \nonumber
\end{equation}
We choose the following VEVs for the link fields $<U_1> = <U_2> =
K_{16 \times 5}$ and $<U_3> = <U_4> = L_{16\times 1}$, where $K_{16
  \times 5}$ is a $16 \times 5$ matrix with elements at $(4,4)$,
$(8,5)$, $(13,1)$, $(14,2)$, and $(15,3)$ being $v/{\sqrt 2}$ and all
other elements $0$, and $L_{16\times 1}$ is a column vector with
$v/{\sqrt 2}$ as the $(12,1)$ element and all other elements $0$.
Similarly to above subsection, we can break the $SO(10)\times {\rm
  flipped} ~SU(5)\times U(1)'$ gauge symmetry down to a diagonal
flipped $\fsu5u1$ gauge symmetry.

Three families of the SM fermions form three ${\bf 16}$ spinor
representations under the first $SO(10)$ while they are singlets and
neutral under the ${\rm flipped} ~SU(5)\times U(1)'$.  We also
introduce a 10-dimensional Higgs field $H_{\bf 10}$, and one pair of
Higgs $\Sigma_{H}$ and ${\overline \Sigma}_{H}$ in the spinor
representation. Moreover, we introduce one pair of the 5-plet fields
$\chi_5$ and ${\overline \chi}_5$, one pair of the 10-plet fields
$\chi_T$ and ${\overline \chi}_T$, one pair of the $U(1)'$ charged
singlets $S_c$ and ${\overline S_c}$, and two singlet fields $S$ and
$S'$.  The quantum numbers for these particles are
\begin{equation}
\label{so10epf44}
\begin{array}{c|cc}
      & SO(10) &  SU(5)\times U(1)'\\
\hline
{\vrule height 15pt depth 5pt width 0pt}
{\bf 16}_i & {\bf 16} & ({\bf 1}, {\bf 0}) \\
H_{\bf 10} & {\bf 10} & ({\bf 1}, {\bf 0}) \\
 \Sigma_{H}      & {\bf 16} & ({\bf 1}, {\bf 0}) \\
 {\overline \Sigma}_{H}  & {\bf {\overline {16}}} & ({\bf 1},  {\bf
   0}) \\
\chi_5 & {\bf 1} & ({\bf {\bar 5}}, {\bf -3}) \\
{\overline \chi}_5 & {\bf 1} & ({\bf { 5}}, {\bf 3}) \\
\chi_T & {\bf 1} & ({\bf { 10}}, {\bf 1}) \\
{\overline \chi}_T & {\bf 1} & ({\bf {\bar 10}}, {\bf -1}) \\
S_s & {\bf 1}  & ({\bf { 1}}, {\bf 5}) \\
{\overline S_s} & {\bf 1}  & ({\bf { 1}}, {\bf -5}) \\
S, S' & {\bf 1}  & ({\bf { 1}}, {\bf 0}) \\
\end{array} \nonumber
\end{equation}

The superpotential is
\begin{eqnarray}
W &=& y_1 \Sigma_{H} U_2 {\overline \chi}_5
+ y_2 {\overline \Sigma}_{H} U_1 \chi_5 + y_3 \Sigma_{H} U_4 {\overline S_s}
+ y_4 {\overline \Sigma}_{H} U_3 S_s
+ y_5 S (\Sigma_{H} {\overline \Sigma}_{H} -M_{\rm FS}^2)
+ \lambda_x \Sigma_{H} H_{\bf 10} \Sigma_{H}
\nonumber\\ &&
+\lambda_y {\overline \Sigma}_{H} H_{\bf 10} {\overline \Sigma}_{H}
+ y_{ij} 16_i H_{\bf 10} 16_j +y_6 S' H_{\bf 10} H_{\bf 10}
+ {1\over M_{\rm Pl}} (y_7 U_1 U_2+ y_8 U_3 U_4) \chi_T {\overline \chi}_T~.~\,
\label{superpotential44}
\end{eqnarray}

The discussions for the proton decay and the gauge coupling
unification are similar to these in the previous subsection. We want
to emphasize that the doublet-triplet splitting problem can be solved
by the missing partner mechanism, and the solution of the proton decay
problem follows.

\section{Conclusions}
\label{sec:conclu}

We have considered the 4-dimensional $N=1$ supersymmetric $SO(10)$
models inspired by the deconstruction of 5-dimensional supersymmetric
orbifold $SO(10)$ models and high dimensional non-supersymmetric
$SO(10)$ models with Wilson line gauge symmetry breaking.  We studied
the $SO(10) \times SO(10)$ models with bi-fundamental link fields
where the $SO(10) \times SO(10)$ gauge symmetry can be broken down to
the PS, $SU(5)\times U(1)$, flipped $SU(5)\times U(1)'$ and the SM
like gauge symmetry.  However, we need to fine-tune the superpotential
to obtain the doublet-triplet splitting.  We proposed two interesting
models: the $SO(10)\times SO(6)\times SO(4)$ model where the gauge
symmetry can be broken down to PS gauge symmetry, and the
$SO(10)\times SO(10)$ model with bi-spinor link fields in which the
gauge symmetry can be broken down to the flipped $\fsu5u1$ gauge
symmetry.  These intermediate gauge symmetries can be further broken
down to the SM gauge symmetry.  In these models, the missing partner
mechanism can be implemented to solve the doublet-triplet splitting
problem, then the higgsino-exchange mediated proton decay is
negligible, and consequently the proton decay is mainly induced by the
dimension-6 operators.  With the GUT scale being at least a few times
$10^{16}$ GeV, the proton lifetime is well above the current
experimental bound.  In addition, we discussed the gauge coupling
unification in these two models: in the $SO(10)\times SO(6)\times
SO(4)$ models, the gauge couplings unify at $3.3\times 10^{16}$ GeV
for a left-right symmetric model and at $5.1\times 10^{16}$ GeV for a
left-right non-symmetric model; and in the $SO(10)\times SO(10)$
model, the gauge couplings unify at about $2.0\times 10^{17}$ GeV.
Furthermore, we briefly commented on the interesting variation models
with gauge groups $SO(10)\times SO(6)$ and $SO(10)\times {\rm
  flipped}~ SU(5)\times U(1)'$ where the proton decay problem can be
solved in the similar manner, and the $SO(10)\times SO(10)$ model with
bi-spinor link fields in which the gauge symmetry can be broken down
to the $SU(5)\times U(1)$ gauge symmetry.

\begin{acknowledgments}
  J. Jiang thanks Csaba Balazs and Cosmas Zachos for helpful
  discussions. The research of C.-S. Huang was supported in part by
  the Natural Science Foundation of China, the research of J. Jiang
  was supported by the U.S.~Department of Energy under Grant
  No.~W-31-109-ENG-38, and the research of T. Li was supported by the
  National Science Foundation under Grant No.~PHY-0070928.

\end{acknowledgments}

\appendix
\section{}
\label{apdx}
The generators and the assignment of the fermions in the {\bf 16} can
be found in Ref.~\cite{Rajpoot:xy}.  We copy the $\sigma \cdot
W_\mu$, and rename it $/\!\!\!\!{A_\mu}$.  The $16\times 16$ matrix
can be re-written into four $8\times 8$ matrices,
\be
\label{eq:gauge16}
/\!\!\!\!{A} =
\left(
\begin{array}{cc}
/\!\!\!\!{A_{11}}&/\!\!\!\!{A_{12}} \\
/\!\!\!\!{A_{21}}&/\!\!\!\!{A_{22}}
\end{array}
\right)\,,
\ee
with
\bea
/\!\!\!\!{A_{11}} &=&
\left(
\begin{array}{cccccccc}
\ld_{1 1}&V_{12}&V_{13}&X_1^0&W_L^{-}& & & \\
V_{12}^*&\ld_{2 2}&V_{23}&X_2^{-}& &W_L^{-}& & \\
V_{13}^*&V_{23}^*&\ld_{3 3}&X_3^{-}& & &W_L^{-}& \\
\overline{X}_1^0&X_2^+&X_3^+&\ld_{4 4}& & &
&W_L^{-}\\
W_L^+& & & &\ld_{5 5}&V_{12}&V_{13}&X_1^0 \\
 &W_L^+& & &V_{12}^*&\ld_{6 6}&V_{23}&X_2^{-} \\
 & &W_L^+& &V_{13}^*&V_{23}^*&\ld_{7 7}&X_3^{-} \\
 & & &W_L^+&\overline{X}_1^0&X_2^+&X_3^+&\ld_{8 8}
\end{array}
\right) \nn \\
/\!\!\!\!{A_{12}} &=& \left(
\begin{array}{cccccccc}
0&A_6^0&-A_5^0&-Y_1^+&0&-Y_6^{-}&Y_5^{-}&-\overline{A}_1^0 \\
-A_6^0&0&A_4^{-}&-\overline{Y}_2^0&Y_6^{-}&0&-Y_4^{--}&-A_2^{-} \\
A_5^0&-A_4^{-}&0&-\overline{Y}_3^0&-Y_5^{-}&Y_4^{--}&0&-A_3^{-} \\
Y_1^+&\overline{Y}_2^0&\overline{Y}_3^0&0&\overline{A}_1^0
&A_2^{-}&A_3^{-}&0\\
0&-A_3^+&A_2^+&-Y_4^{++}&0&Y_3^0&-Y_2^0&-A_4^+ \\
A_3^+&0&A_1^0&-Y_5^+&Y_3^0&0&-Y_1^{-}&-\overline{A}_5^0 \\
-A_2^+&A_1^0&0&-Y_6^+&Y_2^0&-Y_1^{-}&0&-\overline{A}_6^0 \\
Y_4^{++}&Y_5^+&Y_6^+&0&A_4^+&\overline{A}_5^0&\overline{A}_6^0&0
\end{array}
\right) \nn \\
/\!\!\!\!{A_{21}} &=& \left(
\begin{array}{cccccccc}
0&-\overline{A}_6^0&\overline{A}_5^0&Y_1^{-}&0&A_3^{-}&-A_2^{-}&Y_4^{--}\\
\overline{A}_6^0&0&-A_4^+&Y_2^0&-A_3^{-}&0&\overline{A}_1^0&Y_5^{-} \\
-\overline{A}_5^0&A_4^+&0&Y_3^0&A_2^{-}&-\overline{A}_1^0&0&Y_6^{-} \\
-Y_1^{-}&-Y_2^0&-Y_3^0&0&-Y_4^{-}&-Y_5^{-}&-Y_6^{-}&0 \\
0&Y_6^+&-Y_5^+&A_1^0&0&-\overline{Y}_3^0&\overline{Y}_2^0&A_4^{-} \\
-Y_6^+&0&Y_4^{++}&A_2^+&\overline{Y}_3^0&0&-Y_1^+&A_5^0 \\
Y_5^+&-Y_4^{++}&0&A_3^+&-\overline{Y}_2^0&Y_1^+&0&A_6^0 \\
-A_1^0&-A_2^+&-A_3^+&0&-A_4^{-}&-A_5^0&-A_6^0&0
\end{array}
\right) \nn \\
/\!\!\!\!{A_{22}} &=& \left(
\begin{array}{cccccccc}
\ld_{9 9}&-V_{12}^*&-V_{13}^*&-\overline{X}_1^0&W_R^{-}& & & \\
-V_{12}&\ld_{10 10}&-V_{23}^*&-X_2^+& &W_R^{-}& & \\
-V_{13}&-V_{23}&\ld_{11 11}&-X_3^+& & &W_R^{-}& \\
-X_1^0&-X_2^{-}&-X_3^{-}&\ld_{12 12}& & & &W_R^{-} \\
W_R^+& & & &\ld_{13 13}&-V_{12}^*&-V_{13}^*&-\overline{X}_1^0 \\
 &W_R^+& & &-V_{12}&\ld_{14 14}&-V_{23}^*&-X_2^+ \\
 & &W_R^+& &-V_{13}&-V_{23}&\ld_{15 15}&-X_3^+ \\
 & & &W_R^+&-X_1^0&-X_2^-&-X_3^3&\ld_{16 16}
\end{array}
\right) \nn
\eea

The 45 gauge bosons consist of 12 $A$, 6 $X$, 6 $V$, 12 $Y$, 2
charge $W_L$, 2 charge $W_R$ and 16 $\lambda$ which can be rewritten
as 5 independent fields, $V_3$, $V_8$, $V_{15}$, $W_L^0$ and $W_R^0$.

The first family of the SM fermions forms a ${\bf 16}$, \be {\bf
  16}_1=
(u_r,u_g,u_b,\nu_e,d_r,d_g,d_b,e^-,d_r^c,d_g^c,d_b^c,e^+,-u_r^c,-u_g^c,-u_b^c,-\nu_e^c)^T~,
\ee 
similarly for the second and third families.  As the $SO(10)$ is
broken down to $SU(5) \times U(1)$ or flipped $SU(5) \times U(1)'$,
the spinor representation ${\bf 16}$ is decomposed as
\be 
{\bf 16} \to
{\bf 10} + {\bf \bar{5}} + {\bf 1} \ee where \be {\bf 10} =
(Q,U^c,e^+), \quad {\bf \bar{5}} = (D^c,L), \quad {\rm and} \quad {\bf
  1} = \nu^c \ee for breaking to $SU(5) \times U(1)$, and \be {\bf 10}
= (Q,D^c,\nu^c), \quad {\bf \bar{5}} = (U^c,L), \quad {\rm and} \quad
{\bf 1} = e^+ \ee for breaking to flipped $SU(5) \times U(1)'$.

\end{document}